# Determination of Boltzmann constant by gas expansion method

MUSAJ A. PAÇARIZI[1], ARBËR Z. ZEQIRAJ[2], IBRAHIM M. HAMELI[3], ISUF A.TREDHAKU[3], SEFER R. AVDIAJ[3*]

[1]University of Prishtina, Faculty of Natural Sciences and Mathematics, Department of Chemistry, Prishtina 10000, Kosovo

[2]University of Mitrovica, Faculty of Geosciences, Department of Materials and Metallurgy, Mitrovica 40000, Kosovo

[3]University of Prishtina, Faculty of Natural Sciences and Mathematics, Department of Physics, Prishtina 10000, Kosovo

*Corresponding author: sefer.avdiaj@uni-pr.edu

## ABSTRACT

In this paper we present a new method to determine the numerical value of the Boltzmann constant $k$ and its uncertainty. We have used Nitrogen gas in different pressure values in the range 6 kPa – 100 kPa, for three different volumes. In this experiment we have used a simple idea called static expansion method, simple equipment and simple equations for calculations in order to determine the Boltzmann constant. This method is suitable for students of high-school level as well as introductory higher education.

## INTRODUCTION

All basics equations of physics and chemistry have some fundamental invariant quantities called fundamental constants, like $h$-Planck's constant, $c$-universal light speed, G-universal gravity constant, etc. These constants have definite symbols and their numerical values should be measured as accurately as possible. The role of the fundamental constants in science is to describe the relationship between parameters with different dimensions in equations. The numerical value of universal constants tells us the relationship between physical quantities. For example, the value of the universal gravitational constant gives us the information for the intensity of the interaction force between two bodies with certain masses. The values of fundamental constants are supposed to be accepted as a known value in countless experiments during working with students. This is usually the case when they are asked to perform experiments like, the gas-law experiment, or experiments related to semiconductor conductivity. So, it is very interesting for students to measure those fundamental constants in order to understand physical laws much better. In this paper we describe an experiment to measure Boltzmann's constant. The experiment is suitable for the introductory laboratory of chemistry and physics, as the theory of this measurement method is quite simple and is given in many elementary textbooks of chemistry and physics[1,2]. Teaching experimentally the gas laws allows students to discover the relationships between the pressure, volume, temperature, and number of moles (or number of molecules) of a given gas[3] and serves as a foundation to understand kinetic molecular theory.

**Theoretical background of the Boltzmann constant: what is the Boltzmann constant?**
We can study the Boltzmann constant in different phenomena in nature. The Boltzmann constant appears for the first time where it is defined as proportionality constant between the macroscopic entropy of a system and the multiplicity of states for that system:

$$S = k \ln W$$

where S is the system entropy, W is the number of possible microscopic particle configurations for a certain macroscopic state (multiplicity) and k is the Boltzmann constant. Thus, the Boltzmann constant is a proportionality constant which relates microscopic phenomena with macroscopic ones. Hence, each time that a macroscopic effect results from the states of a huge number (of the order of Avogadro's number) of microscopic particles, the mathematical model of that effect, will contain the Boltzmann constant in it.

## MATERIALS AND METHODS

The Boltzmann constant in our case was determined by using ideal gas law[1,2]:

$$PV = NkT \qquad (1)$$

where $P$ – is pressure of the gas, $k$- is Boltzmann's constant, $T$ –is the temperature of the gas in kelvins.

From Eq. (1) the $V$ becomes:

$$V = NTk \cdot \frac{1}{P} \qquad (2)$$

By collecting the data for $P$, $V$ and $T$ the Boltzmann constant $k$, can be calculated if the amount of substance $n$ or number of particles $N$ is known. The problem in this case using this simple equation is how to determine ($N$), the number of molecules (particles) in the closed volume[4]. This is equal to the quotient (3) of the volume V and the standard molar volume $V_m$.

$$N = N_A \frac{V}{V_m} \qquad (3)$$

where $N_A$ is the Avogadro number.

At temperature, $T_0 = 273.15$ K and pressure, $P_0 = 101325$ Pa (standard conditions) the molar volume is $V_m = 0.022414$ m$^3$. The volume $V$, measured at different pressure $P$ and temperature $T$ must first be reduced to these conditions by using the equation:

$$\frac{P_0 V_0}{T_0} = \frac{PV}{T} \qquad (4)$$

The experimental scheme of our measurement method is shown in Figure 1. The idea of measurements is very simple: the gas was expanded in series from a small chamber CH1 to the bigger chamber CH2, and after that to the chamber CH3. Since amount of substance (number of particles) before and after gas expansion should remain the same because the same number of particles (atoms or molecules) was expanded from chamber CH1 to the chamber CH2 after that to the camber CH3. The volume of the gas before expansion is V1, after the first expansion is V1+V2 and after last expansion V1+V2+V3. In the Table 1 are shown the values of the chamber volumes used for this experiment. The chamber volumes are measured by gas expansion method described in reference[5]. The pressure was measured by calibrated vacuum gauge CDG 1000 and the temperature was measured by calibrated thermometer Pt100.

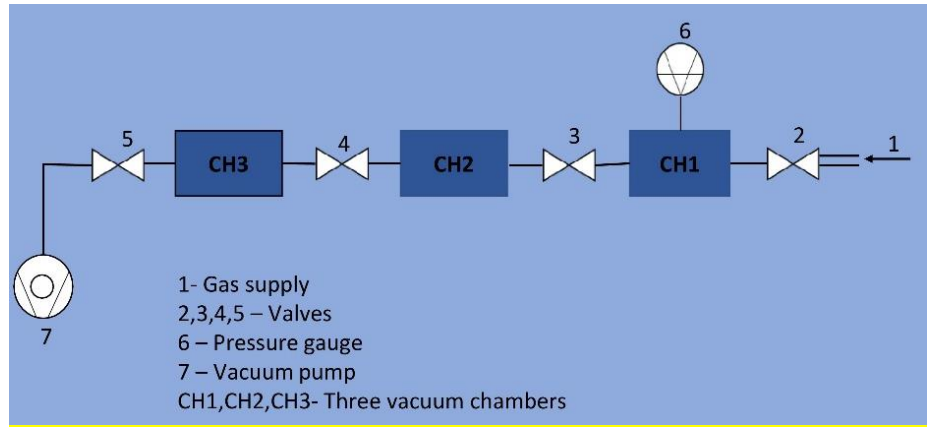

**Figure 1.** Scheme of the experimental setup.

**Table 1. Values of the volumes of three chambers and three different expanded volumes**

|  | CH1 | CH2 | CH3 | CH1+CH2 | CH1+CH2+CH3 |
|---|---|---|---|---|---|
| Volume ($m^3$) | $6.143 \times 10^{-5}$ | $3.369 \times 10^{-4}$ | $3.402 \times 10^{-4}$ | $3.983 \times 10^{-4}$ | $7.385 \times 10^{-4}$ |

## RESULTS AND DISCUSSION

By using Eqs (3) and (4), is possible to determine the number of particles in the chamber for different values of pressure. For example, when the pressure is $P = 400.8$ Torr ($P=53435.458$ Pa), volume $V1 = 6.143 \times 10^{-5}$ $m^3$ and temperature 297.35 K the calculated number of particles is $7.99612 \times 10^{20}$. All measurements were done in the same temperature 297.35 K. As the gas expands it will cool, for this reason, the apparatus was left to return to ambient temperature. The same way was used to calculate the number of particles for different values of initial pressures enclosed in chamber CH1. The measurements were performed by filling CH1 with Nitrogen gas at pressure range 50 Torr – 800 Torr (6666.12 Pa – 106658 Pa). As we mentioned above this number of gas particles were expanded to the chambers CH2 and CH3. So we had two series of expansions. After expansion the pressure is decreased and the volume is increased but the number of particles should remain constant. So each initial pressure was expanded to two following pressures. The measurements were repeated for nine series, each of them with three different pressures and volumes. All the data are included in the appendix.

Since the gas is expanded from one volume to another, it means that the volume was increased while the pressure decreased. The detailed procedure is given in Appendix. From Fig. 2 and Eq.(2) we can conclude that a graph of $V$ against $P^{-1}$ is a straight line with a gradient $NTk$, so that $k$ can be determined using:

$$k = \frac{gradient}{NT} \qquad (5)$$

The equation of the graph given in Fig. 2 tells us there is a small positive intercept on $V$ axis, $1.6 \cdot 10^{-6} m^3$. This intercept is very small compared with smallest experimental value of $V$. There are two possible reasons for this small value of intercept, (a). the gas is considered ideal gas, (b). the possibility of systematic error in $V$. From the gradient in Fig. 2, $gradient = 3.2819 \, m^3 Pa$ and number of molecules $7.99612 \times 10^{20}$ the following value have been calculated for the Boltzmann constant k, according to the equation (5):

$$k = 1.380 \times 10^{-23} JK^{-1}$$

and the relative standard uncertainty for this measurements is:

$$u(k) = 0.78\%$$

The detailed procedure how those values were calculated is given in Appendix.

The value obtained from the CODATA (https://physics.nist.gov/cgi-bin/cuu/Value?k) is:

$$k = 1.380649 \times 10^{-23} JK^{-1}$$

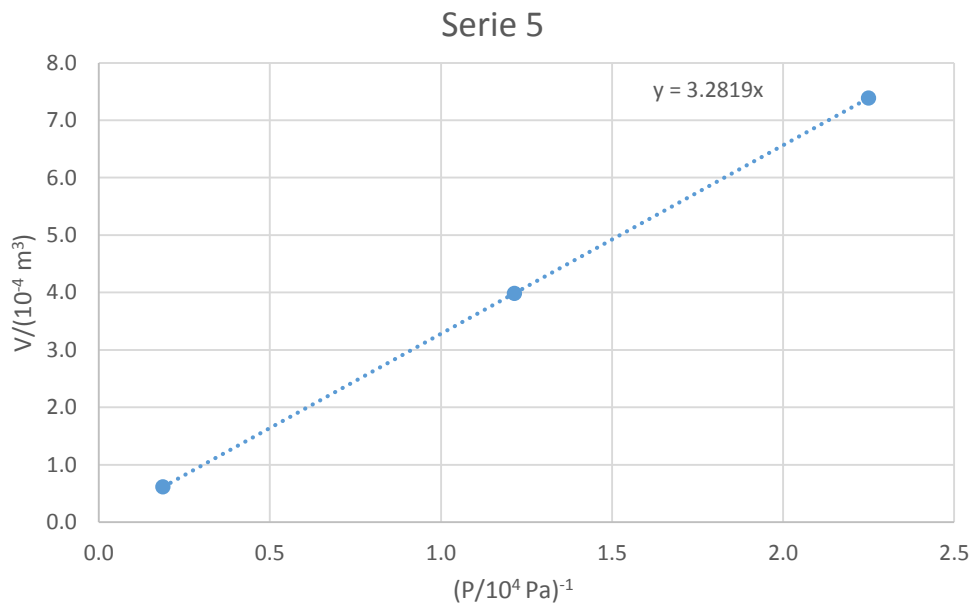

Figure 2. Volume $V$ as function of the reciprocal pressure $1/P$ for constant number of particles.

## CONCLUSIONS

When students determine a fundamental constant they will be removed from the "secret" things to be believed without knowing them. The experiment is using simple equipment and simple equations. These equations are included in every basic book of chemistry and physics. For sure there are much more accurate and precise methods to measure the Boltzmann constant [6,7,8], but they are not appropriate for educational purposes. Also the handling with data in this experiment is very simple.

## ACKNOWNLEDGEMENT

This work was supported mainly by Ministry of Education, Science and Technology of Kosovo. The authors thanks Professor Lars Westerberg from Uppsala University – Sweden for providing some vacuum chambers and vacuum valves.

**Appendix:**

*Table 1. Rough data used for this experiment*

| Measured pressure (Torr) | | | | Volume (m³) | | | Temp.(K) |
|---|---|---|---|---|---|---|---|
| P1 | P2 | P3 | | | | | |
| 50.56 | 7.92 | 4.094 | | V1 | $6.143 \cdot 10^{-5}$ | | 297.35 |
| 100.80 | 15.42 | 8.219 | | V2 | $3.369 \cdot 10^{-4}$ | | |
| 201.20 | 32.20 | 16.330 | | V3 | $3.402 \cdot 10^{-4}$ | | |
| 300.60 | 46.08 | 24.940 | | V1+V2 | $3.983 \cdot 10^{-4}$ | | |
| 400.80 | 61.74 | 33.340 | | V1+V2+V3 | $7.385 \cdot 10^{-4}$ | | |
| 500.40 | 75.76 | 41.910 | | | | | |
| 603.40 | 91.47 | 50.820 | | | | | |
| 701.20 | 109.34 | 59.320 | | | | | |
| 801.40 | 125.01 | 68.420 | | | | | |

**Explanation:**
How the number of molecules in a certain volume was calculated?

Example:

Suppose the gas in enclosed in a chamber being held in temperature $T=297.35$ K and gas pressure is $P = 53435.46$ Pa. The volume of the chamber where the gas is enclosed is $V = 6.143 x 10^{-5} m^3$. How many gas molecules are in this chamber?

Answer:

The volume of one mole gas in standard conditions (temperature $T_\theta = 273,15$ K and pressure $P_\theta = 101325$ Pa) is $V_m = 22.414$ liter ($V_m$=0.022414 m³). One mole of gas in standard conditions has $N_A = 6.022 x 10^{23}$ molecules. If we want to find the number of molecules $N$ in another volume and for different conditions, we can compare

$$\frac{N_A}{V_m} = \frac{N}{V_0} \qquad (1)$$

From Eq. (1) we find,

$$N = N_A \frac{V_0}{V_m} \qquad (2)$$

$V_0$ – is the volume $V$ ($V = 6.143 x 10^{-5} m^3$) reduced to standard conditions.

How to calculate the volume $V_0$ (How to reduce V to standard conditions)?

Simply by comparing standard conditions with new conditions:

$$\frac{P_\theta V_0}{T_\theta} = \frac{PV}{T} \tag{3}$$

So the volume $V_0$ from Eq. (3) should be,

$$V_0 = \frac{PV}{T}\frac{T_\theta}{P_\theta} \tag{4}$$

Inserting the given data yields,

$$V_0 = \frac{53435.46 \text{ Pa} \times 6.143\times10^{-5} \text{m}^3}{297.35 \text{ K}} \frac{273.15 \text{ K}}{101325 \text{ Pa}} = 2.976 \times 10^{-5} \text{m}^3 \tag{5}$$

Substituting for $V_0$ from Eq. (5) into Eq. (2) gives,

$$N = 6.022 \times 10^{23} \frac{2.976\times10^{-5} \text{m}^3}{0.022414 \text{ m}^3} = 7.996 \times 10^{20} \tag{6}$$

This is the calculation for the fifth series of measurements. The eight other measurements for number of molecules are made in the same way.

How did the Boltzmann constant was calculated?

By using ideal gas law:

$$PV = NkT \tag{7}$$

From Eq. (7) the V becomes,

$$V = NTk \cdot \frac{1}{P} \tag{8}$$

From Eq. (7),

$$k = \frac{grad}{NT} \tag{9}$$

where gradient is: $\Delta V/\Delta(P^{-1}) = gradient$. The value of gradient is taken from Figure 2. From the Figure 2 the gradient is: $gradient = 3.2819 \text{ m}^3\text{Pa}$

By using data from Eq. (6), temperature $T$= 297.35 K and the value of gradient we obtain,

$$k = \frac{3.2819 \ m^3\text{Pa}}{7.996 \times 10^{20} \times 297.35 \text{ K}} = 1.380 \times 10^{-23} \text{JK}^{-1}$$

How was calculated the uncertainty in k, $u(k)$?

We will write gradient with letter α.
From Eq.(9) the $k$, depends on,

$$k = k(\alpha, N, T)$$

According to the Guido to the Expression of Uncertainty in Measurements – GUM
https://www.bipm.org/utils/common/documents/jcgm/JCGM_100_2008_E.pdf the
equation for uncertainty in the Boltzmann constant $k$ should be,

$$u(k) = \sqrt{\left(\frac{u(\alpha)}{NT}\right)^2 + \left(\frac{\alpha}{N^2T}u(N)\right)^2 + \left(\frac{\alpha}{NT^2}u(T)\right)^2} \qquad (10)$$

From Eq.(2), $N = N(N_A, V_0, V_m)$, since $N_A$ and $V_m$ are constants, the uncertainties in $N_A$ and $V_m$ are neglected $u(N_A) = 0 \; dhe \; u(V_m) = 0$, then the uncertainty in $N$ should be,

$$u(N) = \frac{N_A}{V_m}u(V_0) \qquad (11)$$

By putting the Eq. (11) to Eq. (10) the uncertainty in k will be ,

$$u(k) = \sqrt{\left(\frac{u(\alpha)}{NT}\right)^2 + \left(\frac{N_A}{V_m}\frac{\alpha}{N^2T}u(V_0)\right)^2 + \left(\frac{\alpha}{NT^2}u(T)\right)^2} \qquad (11)$$

The uncertainty in $V_0$ is calculated from Eq. (4),

$$u(V_0) = \frac{T_\theta}{P_\theta}\sqrt{\left(\frac{V}{T}u(P)\right)^2 + \left(\frac{P}{T}u(V)\right)^2 + \left(\frac{PV}{T^2}u(T)\right)^2} \qquad (12)$$

The uncertainty of measured pressure, volume and temperature were:
$u(p) = 0.2\%$, $u(V) = 0.75\%$ and $u(T) = 0.05$ K. The uncertainty of pressure and temperature were taken from calibration certificates of the instruments whereas the uncertainty of the volume was taken from reference 5.

The uncertainty in $V_0$ should be,

$$u(V_0) = \frac{273.15 \text{ K}}{101325 \text{ Pa}} \times$$
$$\sqrt{\left(\frac{6.143 \times 10^{-5}\text{m}^3}{297.35 \text{ K}} \times 106.87 \text{ Pa}\right)^2 + \left(\frac{53435.46 \text{ Pa}}{297.35 \text{ K}} \times 4.61 \times 10^{-7}\text{m}^3\right)^2 + \left(\frac{53435.46 \text{ Pa} \times 6.143 \times 10^{-5}\text{m}^3}{(297.35 \text{ K})^2} \times 0.05 \text{ K}\right)^2}$$

$$u(V_0) = 2.31 \times 10^{-7} \text{m}^3$$

From Figure 2, the uncertainty in the gradient was calculated to be, $u(\alpha) = 0.0012$ m³Pa. Using the Eq. (11) and the data for α, $u(\alpha)$, $N$, $T$, $N_A$, $V_m$, $u(V_0)$, $u(T)$, the uncertainty in $k$ should be,

$$u(k) = 1.06 \times 10^{-25} \text{JK}^{-1}$$

The relative uncertainty of $k$,

$$\frac{u(k)}{k} = 0.78 \text{ \%}$$

The other values of $k$ and $u(k)$ can be evaluated through similar calculations.

Figure 3. Volume vs reciprocal pressure for all series of measurements